\documentclass[english,aps,aps,pra,twocolumn,preprintnumbers,amsmath,amssymb]{revtex4-1}
\usepackage[T1]{fontenc}
\usepackage[latin9]{inputenc}
\usepackage{array}
\usepackage{float}
\usepackage{multirow}
\usepackage{amsmath}
\usepackage{amssymb}
\usepackage{graphicx}
\usepackage{bbm}
\usepackage{dsfont}

\makeatletter

\providecommand{\tabularnewline}{\\}

\makeatother

\usepackage{babel}
\begin{document}

\title{Estimating ergodization time of a chaotic many-particle system from
a time reversal of equilibrium noise}

\author{Andrei E. Tarkhov$^{1}$ and Boris V. Fine$^{1,2}$}

\affiliation{$^{1}$Skolkovo Institute of Science and Technology, Skolkovo Innovation
Center, Novaya~street~100, Skolkovo~143025, Russia}

\affiliation{$^{2}$Institute for Theoretical Physics, University of Heidelberg,
Philosophenweg 12, 69120 Heidelberg, Germany}

\date{\today}
\begin{abstract}
We propose a method of estimating ergodization time of a chaotic many-particle
system by monitoring equilibrium noise before and after time reversal
of dynamics~(Loschmidt echo). The ergodization time is defined as
the characteristic time required to extract the largest Lyapunov exponent
from a system's dynamics. We validate the method by numerical simulation
of an array of coupled Bose-Einstein condensates in the regime describable
by the discrete Gross-Pitaevskii equation. The quantity of interest
for the method is a counterpart of out-of-time-order correlators~(OTOCs)
in the quantum regime.
\end{abstract}
\maketitle

\section{Introduction}

Quantitative characterization of ergodicity in many-particle systems
is a long-standing challenge for the foundations of statistical physics,
which dates back to the Poincar\'e recurrence theorem~\cite{poincare2003three}
and Zermelo's paradox~\cite{zermelo2003theorem}. It was already
pointed out by Boltzmann~\cite{Lebowitz199332,boltzmann1896entgegnung}
and since then became fairly obvious for the practitioners in the
field~\cite{birkhoff1931proof,khinchin1949mathematical} that the
ergodization time of many-particle systems, defined as the Poincar\'e
recurrence time, is impractically long to be observable on experimental
timescales. Instead, it is common to call many-particle systems ``ergodic'',
when they establish the Boltzmann-Gibbs equilibrium on an experimentally
observable timescale. But even with such a concept in mind, it still
remains a challenge to define the corresponding ergodization time
and to measure this time experimentally.

In this paper, we define the ergodization time of a chaotic system
as the characteristic time one needs to monitor the system in order
to extract its primary chaotic parameter, namely, the largest Lyapunov
exponent, which characterizes the sensitivity of a system to infinitesimal
perturbations, the so-called ``butterfly effect''. The advantage
of this definition is that it is unbiased in the sense of not being
coupled to any particular system's coordinate. Our goal is to theoretically
propose and numerically test a method, which can be used to experimentally
determine whether the system is ergodic or not, and if it is, then
to extract the system's ergodization time. The method is based on
monitoring the equilibrium noise of the system. It involves the time
reversal of system's dynamics \textemdash{} the so-called ``Loschmidt
echo''~\cite{gorin2006dynamics,goussev2012loschmidt}. 

Various aspects of this work are relevant to the previous investigations
of lattice gauge models~\cite{bolte2000ergodic,fulop2001towards,biro2004chaotic,kunihiro2010chaotic,iida2013entropy}
and spin lattice models~\cite{de2012largest,de2013lyapunov,fine2014absence,elsayed2015sensitivity}.
We also note that our method involves the classical counterpart of
out-of-time-order quantum correlators (OTOCs)~\cite{larkin1969quasiclassical}
that have been actively investigated in recent years in the context
of quantum thermalization~\cite{maldacena2016bound,bohrdt2016scrambling,syzranov2017out,rozenbaum2017lyapunov,garttner2017measuring}
and many-body localization problems~\cite{he2017characterizing,swingle2017slow,slagle2017out,fan2017out,serbyn2017loschmidt,huang2017out}.
The relation between our results and OTOCs is to be discussed at the
end of this paper.

\begin{figure}
\includegraphics[width=1\columnwidth]{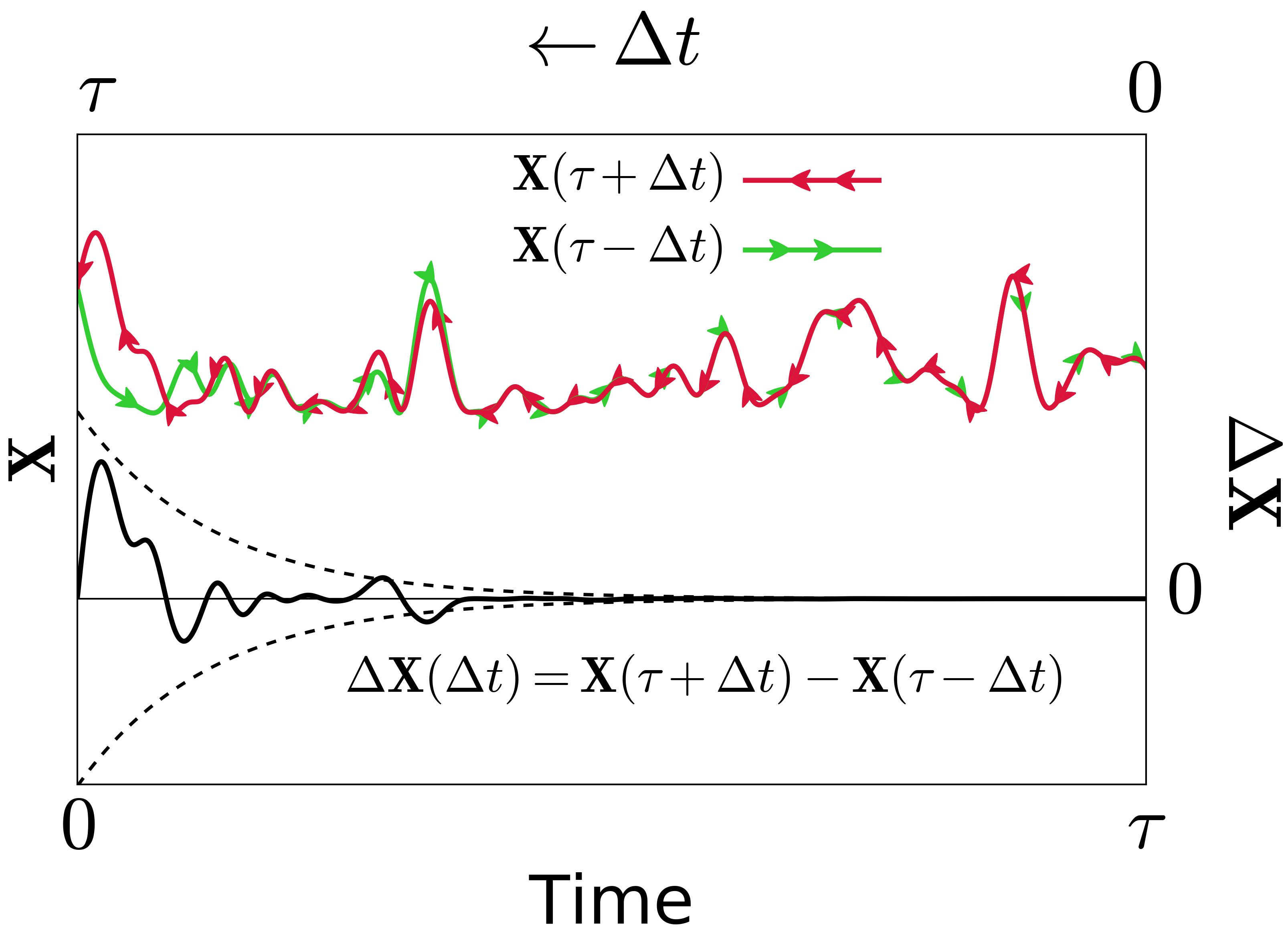}\caption{(Color online) Sketch of a slightly imperfect noise reversal. Equilibrium
noise of an observable $\mathbf{X}$ before and after an imperfect
time reversal of a system's dynamics at time $\tau$ is denoted, respectively,
as $\mathbf{X}(\tau-\Delta t)$ (green line) and $\mathbf{X}(\tau+\Delta t)$
(red line), where $\Delta t=\left|t-\tau\right|$. In order to facilitate
visual comparison, ``Time'' on the horizontal axis represents $t$
before the time reversal and $2\tau-t$ after the time reversal. The
difference between the direct and the reversed noise $\Delta\mathbf{X}(\Delta t)=\mathbf{X}(\tau+\Delta t)-\mathbf{X}(\tau-\Delta t)$
(thick black line) fluctuates around $0$, while its amplitude grows,
on average, exponentially as a function $\Delta t$ with a rate equal
to the largest Lyapunov exponent $\lambda_{\max}$. The exponentially
growing envelope of $\Delta\mathbf{X}(\Delta t)$ is represented by
dashed lines.\label{fig:Sketch_of_the_method}}
\end{figure}

\section{Outline of the method}

In Fig.~\ref{fig:Sketch_of_the_method}, we outline the method. It
consists of the following steps. 

(i) Measuring equilibrium noise of observable $\mathbf{X}$ before
and after slightly imperfect time reversal. The noise is to be denoted
as $\mathbf{X}(\tau-\Delta t)$ and $\mathbf{X}(\tau+\Delta t)$,
where $\tau$ is the time of the dynamics' reversal, and $\Delta t=\left|t-\tau\right|$. 

(ii) Calculating the difference $\Delta\mathbf{X}(\Delta t)\equiv\mathbf{X}(\tau+\Delta t)-\mathbf{X}(\tau-\Delta t)$. 

(iii) Repeating the procedure for an ensemble of randomly chosen initial
conditions on an energy shell. 

(iv) Calculating two kinds of ensemble averages $\left\langle \ln\left|\Delta\mathbf{X}(\Delta t)\right|\right\rangle $
and $\ln\left\langle \left|\Delta\mathbf{X}(\Delta t)\right|\right\rangle $.
For $\Delta t\to\infty$, the former average approaches $\lambda_{\max}\Delta t$,
while the latter one approaches $\Lambda\Delta t$, where $\lambda_{\max}$
is the largest Lyapunov exponent, and $\Lambda$ is a parameter to
be discussed later. 

(v) Extracting the ergodization time $\tau_{erg}$, which, as we show
below, is proportional to the difference between $\Lambda$ and $\lambda_{\max}$.

\begin{figure}
\includegraphics[width=1\columnwidth]{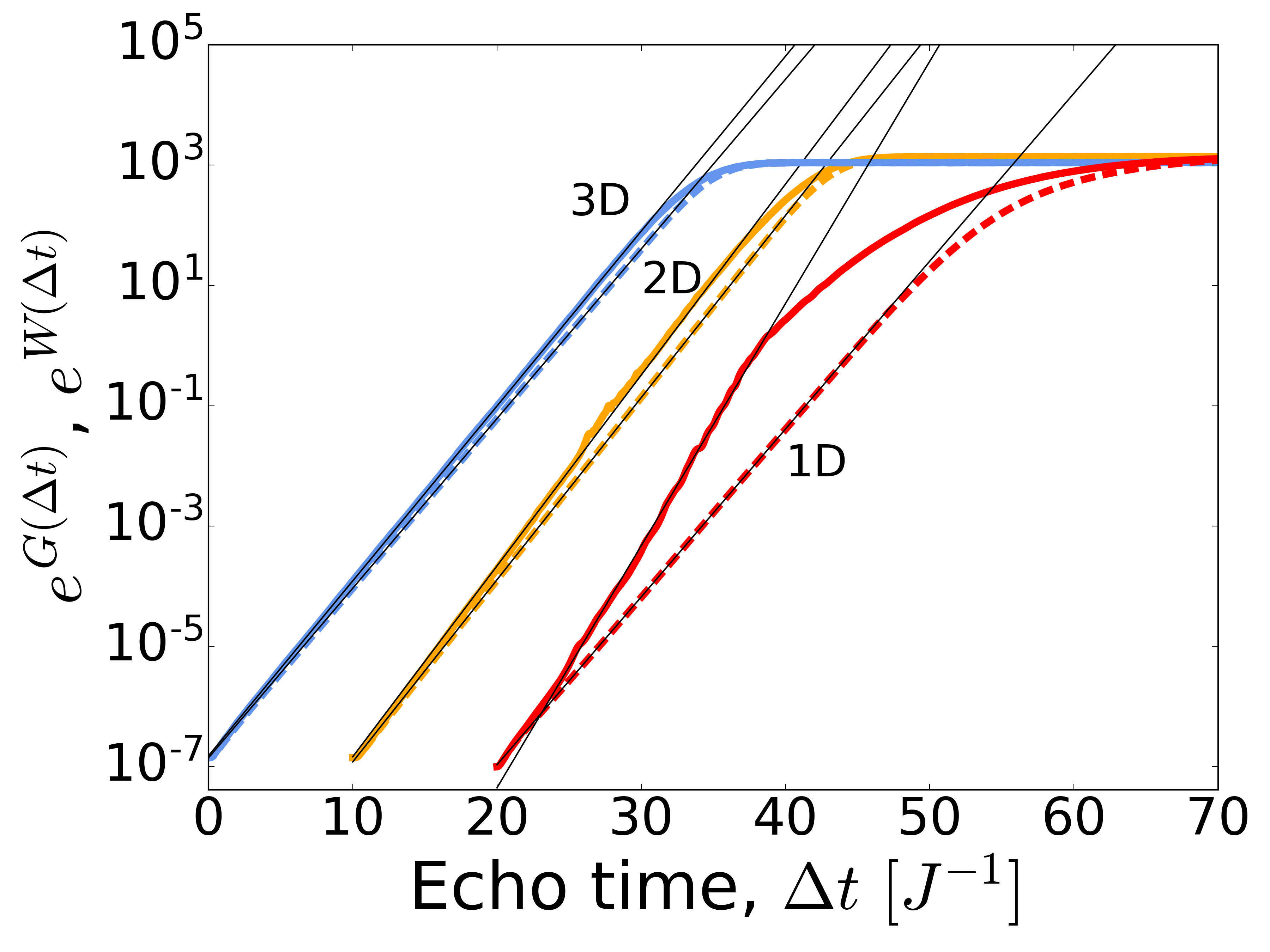}

\caption{(Color online) Loschmidt echo responses $G(\Delta t)$ defined according
to Eq.~(\ref{eq:G_definition}) (dashed lines), and $W(\Delta t)$
defined by Eq.~(\ref{eq:W_definition}) (solid lines) for: a three-dimensional
$4\times4\times4$ cubic lattice (3D, light blue); a two-dimensional
$10\times10$ square lattice (2D, orange, shifted to the right by
$10$); a one-dimensional chain with $100$ sites~(1D, red, shifted
to the right by $20$). Thin black lines are linear fits from which
$\lambda_{\max}$ and $\Lambda$, listed in Table~\ref{tab:Comparison-of-all_the_resutls},
were extracted.\label{fig:Loschmidt_echo_for_dN_vs_LLE}}
\end{figure}

\section{Model system}

The method is generally applicable to systems where time-reversal
of the dynamics can be practically implemented. Here, we illustrate
it for one-, two- and three-dimensional arrays of coupled Bose-Einstein
condensates~(BECs) in the regime describable by the discrete Gross-Pitaevskii
equation~(DGPE):

\begin{equation}
i\frac{d\psi_{j}}{dt}=-J\sum_{k}^{\text{NN}(j)}\psi_{k}+\beta\left|\psi_{j}\right|^{2}\psi_{j},\label{eq:dynamical_equations}
\end{equation}
where $\psi_{j}$ is the complex order-parameter, describing the condensate
at site $j=1\ldots N$, $J$ is the hopping parameter, and $\beta$
is the nonlinear on-site interactions parameter, respectively. The
summation over $k=1\ldots N_{\text{nn}}$ extends over the nearest-neighbors
$\text{NN}(j)$ of site $j$. The DGPE generates conservative dynamics
corresponding to the Hamiltonian 
\begin{equation}
{\cal H}=-J\sum_{\left\langle i,j\right\rangle }\psi_{i}^{*}\psi_{j}+\frac{\beta}{2}\sum_{i}\left|\psi_{i}\right|^{4}.\label{eq:Hamiltonian}
\end{equation}

As a measurable quantity of interest we have chosen a set of on-site
occupations $\mathbf{X}(t)=\{n_{1},n_{2},\ldots,n_{N}\}$, where $n_{i}\equiv\left|\psi_{i}^{2}\right|$
.

\section{Ergodization time}

\subsection{Definitions of Lyapunov exponents and ergodization time}

The largest Lyapunov exponent is defined as $\lambda_{\max}\equiv\frac{1}{t}\lim_{t\to\infty,D(0)\to0}\left(\ln\left|\frac{D(t)}{D(0)}\right|\right),$
where $D(t)=\left|\mathbf{\delta R}(t)\right|$ is the distance between
the two phase-space trajectories: the reference trajectory $\mathbf{R}_{1}(t)$
and the slightly perturbed one $\mathbf{R}_{2}(t)=\mathbf{R}_{1}(t)+\mathbf{\delta R}(t)$~\cite{wimberger2014nonlinear}.

The ratio $\ln\left|\frac{D(t)}{D(0)}\right|$ fluctuates in time
as the reference trajectory $\mathbf{R}_{1}(t)$ explores the energy
shell. We define instantaneous local stretching rates as $\lambda(t)=\frac{d}{dt}\ln\left|\frac{D(t)}{D(0)}\right|$.
The largest Lyapunov exponent is the average of local stretching rates
over a sufficiently long time: $\lambda_{\max}=\overline{\lambda(t)}$.
We denote fluctuations of the local stretching rates by $\delta\lambda\left(t\right)\equiv\lambda\left(t\right)-\lambda_{\max}$,
and their autocorrelator by 

\begin{equation}
\varphi(t)\equiv\left\langle \delta\lambda(t)\delta\lambda(0)\right\rangle .\label{eq:autocorrelator_definition}
\end{equation}
We propose to use the convergence of $\overline{\lambda(t)}$ as an
indicator of ergodization, and define the ergodization time as
\begin{equation}
\tau_{erg}\equiv\frac{1}{\left\langle \delta\lambda^{2}\right\rangle }\int_{0}^{\infty}\varphi(t)dt.\label{eq:tau_erg_from_autocorrelation}
\end{equation}
In numerical simulations, $\lambda_{\max}$ and $\varphi(t)$ can
be obtained from the direct calculations of $\mathbf{R}_{1}(t)$ and
$\mathbf{R}_{2}(t)$. However, such an approach is impractical experimentally,
because it requires tracking all phase-space coordinates of the system.
An alternative, more practical approach was proposed in Refs.~\cite{fine2014absence,tarkhov2017extracting}.
That approach is based on monitoring the effect of Loschmidt echo
on equilibrium noise of almost any observable (see Appendix~\ref{sec:from_phase_space_dist_to_G_and_W}).

\subsection{Ergodization time from Loschmidt echoes}

In the present setting, the Loschmidt echo is implemented by reversing
the sign of Hamiltonian~(\ref{eq:Hamiltonian}) at time $\tau$,
and simultaneously perturbing the state vector, $\psi_{i}(\tau+0)=\psi_{i}(\tau-0)+\delta\psi_{i}$,
where $\delta\psi_{i}$ is a very small random perturbation. We track
the equilibrium noise of the on-site occupations $\{n_{i}(t)\}$ before
and after the time reversal, and introduce the deviation between the
reversed and direct dynamics of the on-site occupations, $\Delta n_{i}(\Delta t)\equiv n_{i}(\tau+\Delta t)-n_{i}(\tau-\Delta t)$.
As sketched in Fig.~\ref{fig:Sketch_of_the_method}, the deviations
$\Delta n_{i}(\Delta t)$ fluctuate with an exponentially-growing
envelope.

We introduce two ensemble averages of $\Delta n_{i}(\Delta t)$:

\begin{equation}
G(\Delta t)\equiv\left\langle \ln\sqrt{\sum_{i=1}^{N}\left[\Delta n_{i}(\Delta t)\right]^{2}}\right\rangle \xrightarrow[\Delta t\to\infty]{}\lambda_{\max}\Delta t\label{eq:G_definition}
\end{equation}
and 

\begin{equation}
W(\Delta t)\equiv\ln\left\langle \sqrt{\sum_{i=1}^{N}\left[\Delta n_{i}(\Delta t)\right]^{2}}\right\rangle \xrightarrow[\Delta t\to\infty]{}\Lambda\Delta t,\label{eq:W_definition}
\end{equation}
where $\Lambda\equiv\frac{1}{t}\ln\left\langle \exp\int_{0}^{t}\lambda(t')dt'\right\rangle $.

\begin{figure}[t]
\begin{raggedright}
\includegraphics[width=1\columnwidth]{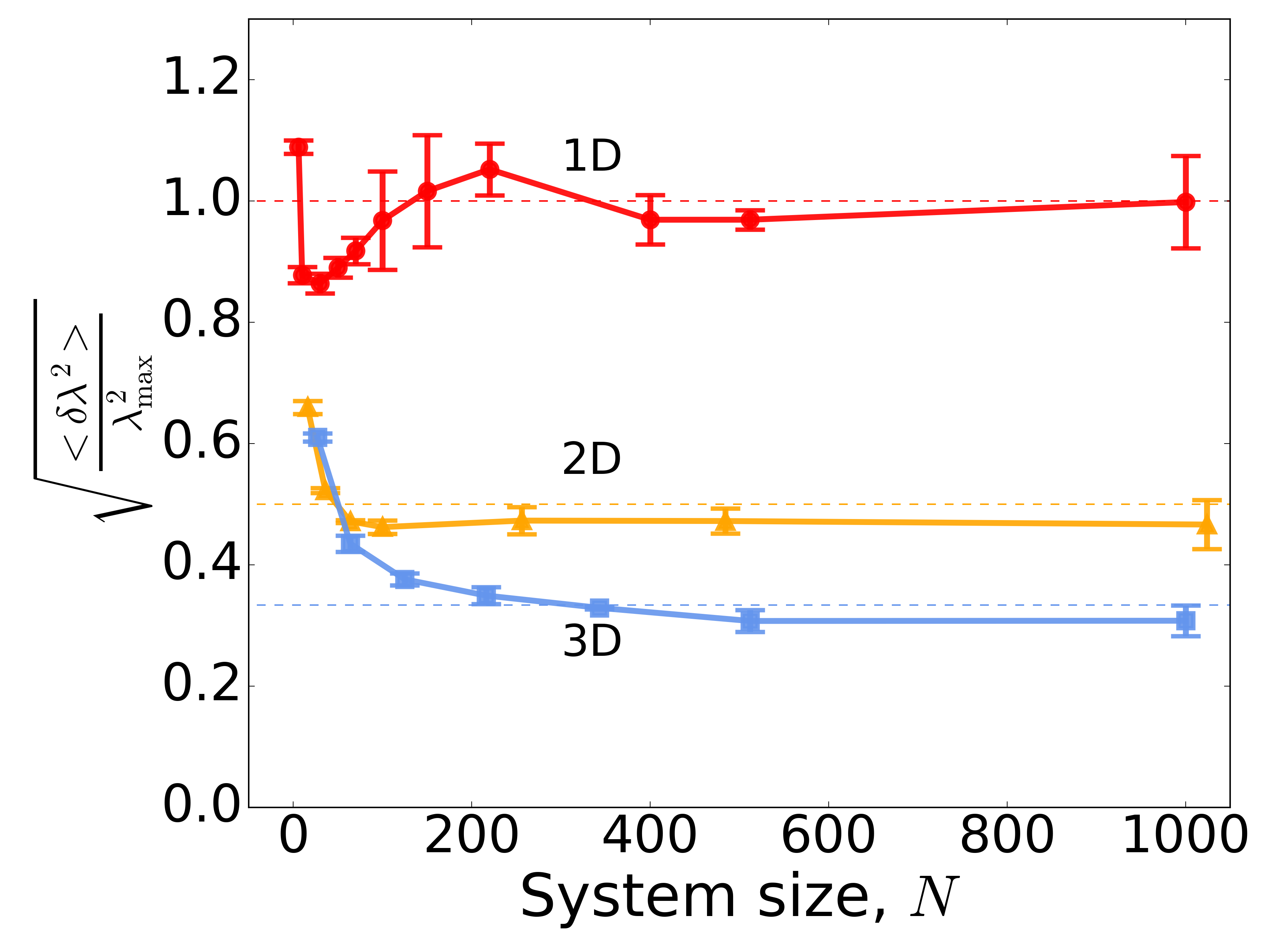}
\par\end{raggedright}
\caption{(Color online) Numerical test of empirical estimate~(\ref{eq:approximate_variance_LSR})
for $\left\langle \delta\lambda^{2}\right\rangle $. The dependence
of $\sqrt{\left\langle \delta\lambda^{2}\right\rangle /\lambda_{\max}^{2}}$
on the number of lattice sites $N$ for one-~(red circles), two-~(orange
triangles) and three-dimensional~(light blue squares) lattices. The
dashed lines are plotted at the levels of $2/N_{\text{nn}}$. \label{fig:System_size_dependence}}
\end{figure}

The limit~(\ref{eq:G_definition}) was verified recently in Ref.~\cite{tarkhov2017extracting}.
Now, we concentrate on relation~(\ref{eq:W_definition}). The reason
for the difference between parameter $\Lambda$~(sometimes referred
to as the generalized maximum Lyapunov exponent~\cite{fujisaka1983statistical,benzi1985characterisation,benzi1999characterization,akimoto2015generalized})
and $\lambda_{\max}$ is the different order of operations of taking
logarithm and ensemble averaging. This difference is controlled by
the amplitude and the correlation time of fluctuations $\delta\lambda(t)$.
In order to demonstrate this, we first note that 
\begin{equation}
\Lambda-\lambda_{\max}=\frac{1}{t}\ln\left\langle e^{\int_{0}^{t}\delta\lambda(t')dt'}\right\rangle .\label{eq:difference_Lam_lam}
\end{equation}
The average on the right-hand side can be calculated analytically
on the basis of the assumption that variable $\int_{0}^{t}\delta\lambda(t')dt'$
is Gaussian, which gives~(see Appendix~\ref{sec:Calculation-of-the_correction}):

\begin{equation}
\left\langle e^{\int_{0}^{t}\delta\lambda(t')dt'}\right\rangle =e^{t\int_{0}^{\infty}\varphi(t')dt'}.\label{eq:equation_for_derivation_in_SUPP}
\end{equation}
Using this relation together with Eq.~(\ref{eq:tau_erg_from_autocorrelation}),
we obtain $\Lambda-\lambda_{\max}=\int_{0}^{\infty}\varphi(t')dt'\equiv\left\langle \delta\lambda^{2}\right\rangle \tau_{erg}$.
Therefore, the ergodization time can be expressed as

\begin{equation}
\tau_{erg}=\frac{\Lambda-\lambda_{\max}}{\left\langle \delta\lambda^{2}\right\rangle }.\label{eq:ergodization_time_estimate_1}
\end{equation}

\subsection{Extracting the ergodization time of DGPE lattices by measuring observables
only}

The experimental use of Eq.~(\ref{eq:ergodization_time_estimate_1})
requires determining $\lambda_{\max}$ and $\Lambda$ from Eqs.~(\ref{eq:G_definition})
and (\ref{eq:W_definition}) and, in addition, the knowledge of $\left\langle \delta\lambda^{2}\right\rangle $.
While there might be ways of extracting $\left\langle \delta\lambda^{2}\right\rangle $
from experimental time-series, here we resort to an empirical estimate
\begin{equation}
\left\langle \delta\lambda^{2}\right\rangle \approx\frac{4\lambda_{\max}^{2}}{N_{\text{nn}}^{2}},\label{eq:approximate_variance_LSR}
\end{equation}
where $N_{\text{nn}}^{2}$ is the number of nearest neighbors for
a lattice site. In Fig.~\ref{fig:System_size_dependence}, we substantiate
the estimate~(\ref{eq:approximate_variance_LSR}) on the basis of
our direct numerical simulations. Why this approximation works so
well for the DGPE on large lattices and whether it works for a more
general class of systems needs further investigation. A possible explanation
of Eq.~(\ref{eq:approximate_variance_LSR}) is that, in our simulations,
the Lyapunov eigenvector corresponding to $\lambda_{\max}$ is usually
localized at only a handful of sites, which is consistent with other
observations of wandering localization of Lyapunov eigenvectors~\cite{falcioni1991ergodic,pikovsky1998dynamic,ruffo1999lyapunov,pikovsky2001dynamic,posch2002lyapunov,taniguchi2003localized,bosetti2010covariant,kuptsov2014predictable}.

The estimate~(\ref{eq:approximate_variance_LSR}) leads to the following
approximation for the ergodization time

\begin{equation}
\tau_{erg}\approx\frac{\Lambda-\lambda_{\max}}{4\lambda_{\max}^{2}}N_{\text{nn}}^{2}.\label{eq:ergodization_time_rough_estimate_2}
\end{equation}

\section{Criterion of ergodicity\label{sec:Criterion-of-ergodicity}}

When the ergodicity of a system is about to break down, one obvious
indicator of this is an anomalously large value of the ergodization
time given by Eq.~(\ref{eq:ergodization_time_estimate_1}). One may
wonder, however, whether the Loschmidt echo response contains other
signatures of broken ergodicity. In an ergodic regime, the distribution
of $\ln\left|\Delta\mathbf{X}(\Delta t)\right|$ should be Gaussian~(see
Appendix~\ref{sec:Calculation-of-the_correction}), and its variance
$\sigma_{G}^{2}(\Delta t)\equiv\left\langle \ln^{2}\left|\Delta\mathbf{X}(\Delta t)\right|\right\rangle -G^{2}(\Delta t)$
is supposed to grow linearly in time:

\begin{equation}
\sigma_{G}^{2}(\Delta t)\xrightarrow[\Delta t\to\infty]{}2(\Lambda-\lambda_{\max})\Delta t.\label{eq:variance_of_G}
\end{equation}
In the opposite case of a non-ergodic regime, the averages in $G(\Delta t)$
and $W(\Delta t)$ converge poorly, which in turn leads to a non-Gaussian
distribution for individual realizations of $\ln\left|\Delta\mathbf{X}(\Delta t)\right|$~\cite{vulpiani2010chaos},
accompanied by a deviation from the linear growth of $\sigma_{G}^{2}(\Delta t)$
given by Eq.~(\ref{eq:variance_of_G}). Thus, relation~(\ref{eq:variance_of_G})
can be used for an experimentally feasible test of ergodization. 

\section{Numerical tests}

For illustration, we chose three model systems: a one-dimensional
chain with $N=100$ sites, a two-dimensional square lattice with $N=10\times10$
sites and a three-dimensional cubic lattice with $N=4\times4\times4$.
We used $J=1$, $\beta=0.01$. The initial conditions corresponded
to the total energy $E_{total}=100N$ and the number of particles
$N_{p}\equiv\sum_{i}\left|\psi_{i}\right|^{2}=100N$, so that the
particles were distributed equally among all lattice sites $n_{i}(0)\equiv\left|\psi_{i}(0)\right|^{2}=100$
with random phases. The perturbation to the state vector at the moment
of time reversal was $\psi_{i}(\tau+0)=\psi_{i}(\tau-0)+\delta\psi_{i}$,
where $\delta\psi_{i}$ is a random vector in the phase space subject
to the constraint $\sqrt{\sum_{i}\left|\delta\psi_{i}\right|^{2}}=10^{-8}$.

In order to test the relation~(\ref{eq:ergodization_time_estimate_1}),
we calculated the two averages of Loschmidt echoes $G(\Delta t)$
and $W(\Delta t)$ for one-, two- and three-dimensional DGPE lattices.
The results are presented in Fig.~\ref{fig:Loschmidt_echo_for_dN_vs_LLE}.
The values of the characteristic exponents $\lambda_{\max}$ and $\Lambda$
extracted in each case are listed in Table~\ref{tab:Comparison-of-all_the_resutls}.
We also collected long enough time-series of local stretching rates
$\lambda(t)$, then calculated the autocorrelation function $\varphi(t)$
and extracted $\left\langle \delta\lambda^{2}\right\rangle $ and
$\tau_{erg}$.

\begin{table*}[t]
\begin{tabular}{|l|c|c|c|c|c|c|c|c|}
\hline 
\multirow{2}{*}{Lattice} & \multirow{2}{*}{~$N_{\text{nn}}$~} & \multirow{2}{*}{~~~$\lambda_{\max}$~~~} & \multirow{2}{*}{~~~$\Lambda$~~~} & \multicolumn{2}{c|}{$\left\langle \delta\lambda^{2}\right\rangle $} & \multicolumn{3}{c|}{$\tau_{erg}$}\tabularnewline
\cline{5-9} 
 &  &  &  & ~~~Eq.~(\ref{eq:tau_erg_from_autocorrelation})~~~ & ~~~Eq.~(\ref{eq:approximate_variance_LSR})~~~ & ~~~Eq.~(\ref{eq:tau_erg_from_autocorrelation}) & ~~~Eq.(\ref{eq:ergodization_time_estimate_1})~~~ & ~~~Eq.~(\ref{eq:ergodization_time_rough_estimate_2})~~~\tabularnewline
\hline 
1D, $N=100$ & $2$ & $0.643\pm0.001$ & $0.927\pm0.009$ & $0.362\pm0.001$ & $0.413\pm0.001$ & $0.66\pm0.05$ & $0.78\pm0.03$ & $0.69\pm0.02$\tabularnewline
2D, $N=10\times10$ & $4$ & $0.698\pm0.001$ & $0.731\pm0.004$ & $0.104\pm0.001$ & $0.122\pm0.001$ & $0.32\pm0.02$ & $0.32\pm0.04$ & $0.27\pm0.03$\tabularnewline
3D, $N=4\times4\times4$~ & $6$ & $0.650\pm0.001$ & $0.670\pm0.001$ & $0.080\pm0.001$ & $0.047\pm0.001$ & $0.26\pm0.02$ & $0.25\pm0.02$ & $0.43\pm0.03$\tabularnewline
\hline 
\end{tabular}

\noindent \caption{Summary of numerical tests of relations~(\ref{eq:ergodization_time_estimate_1})
and (\ref{eq:ergodization_time_rough_estimate_2}): $\lambda_{\max}$
and $\Lambda$ are extracted from Fig.~\ref{fig:Loschmidt_echo_for_dN_vs_LLE};
$\left\langle \delta\lambda^{2}\right\rangle $ is extracted either
directly from a time-series of local stretching rates according to
Eq.~(\ref{eq:tau_erg_from_autocorrelation}) or from emprical estimate~(\ref{eq:approximate_variance_LSR});
the three values of $\tau_{erg}$ are obtained on the basis of the
definition~(\ref{eq:tau_erg_from_autocorrelation}), from the Loschmidt
echo relation~(\ref{eq:ergodization_time_estimate_1}), and from
the approximate relation~(\ref{eq:ergodization_time_rough_estimate_2}).\label{tab:Comparison-of-all_the_resutls}}
\end{table*}

Table~\ref{tab:Comparison-of-all_the_resutls} compares three values
of the ergodization time: the one calculated on the basis of the definition~(\ref{eq:tau_erg_from_autocorrelation}),
the one given by Eq.(\ref{eq:ergodization_time_estimate_1}) and the
one given by the approximation~(\ref{eq:ergodization_time_rough_estimate_2}).
In Eq.(\ref{eq:ergodization_time_estimate_1}), we used the directly
calculated value of $\left\langle \delta\lambda^{2}\right\rangle $.

\begin{figure}[H]
\begin{raggedright}
\includegraphics[width=1\columnwidth]{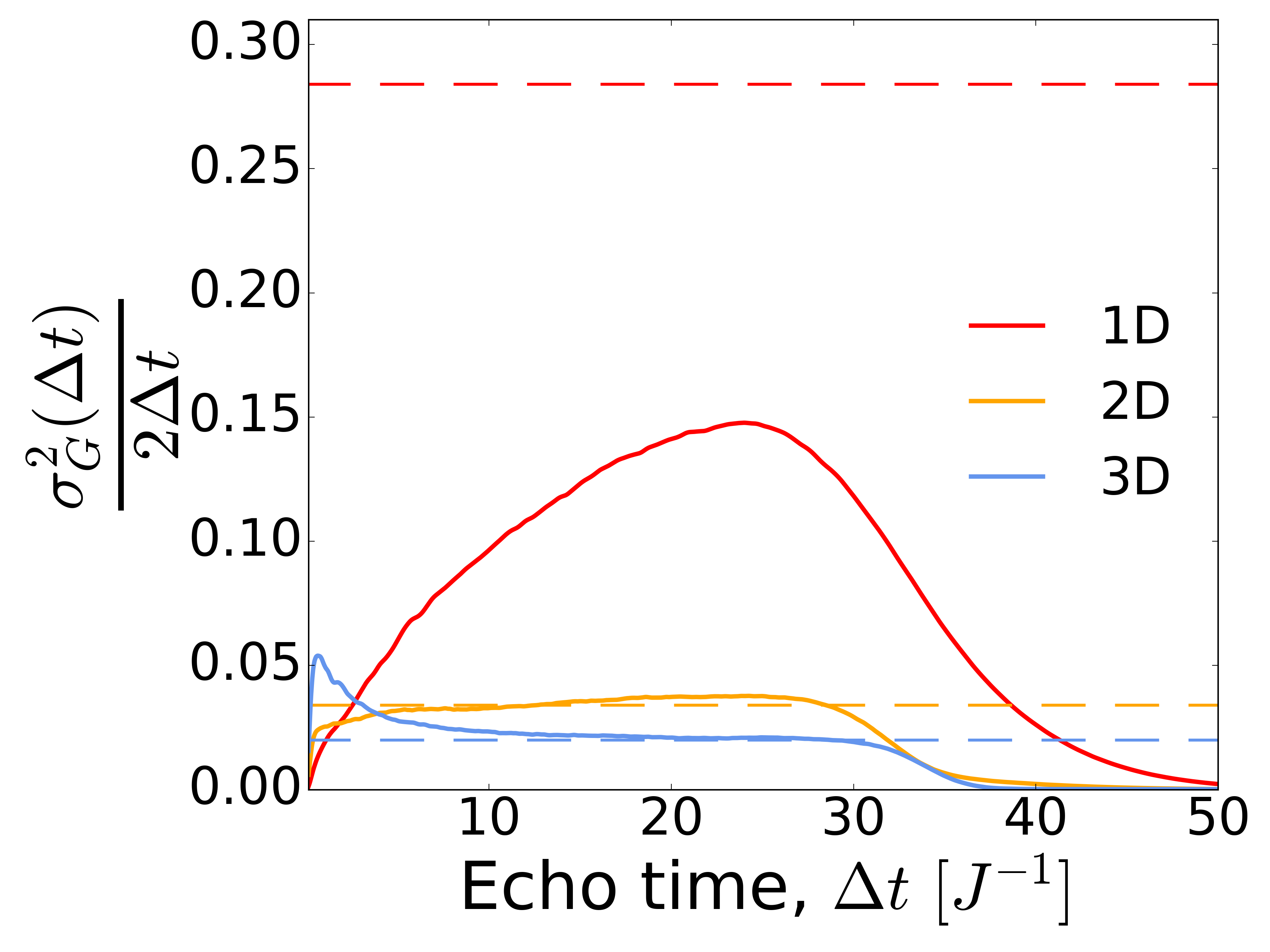}
\par\end{raggedright}
\caption{(Color online) Ergodicity tests. The dependence of the ratio $\frac{\sigma_{G}^{2}(\Delta t)}{2\Delta t}$
on the echo time $\Delta t$ for a 1D chain of $100$ sites~(red),
a 2D square lattice $10\times10$~(orange), a 3D cubic lattice $4\times4\times4$~(light
blue). The dashed lines are plotted at the levels $\Lambda-\lambda_{\max}$
corresponding to the plateaux expected for ergodizing systems. These
plots imply that the 2D and 3D lattices are ergodized on the timescale
of our simulations, while the 1D lattice is not. \label{fig:SUPP_X_sqr_plateau}}
\end{figure}

For two- and three-dimensional lattices, the values of the ergodization
time obtained from Eqs.~(\ref{eq:tau_erg_from_autocorrelation})
and (\ref{eq:ergodization_time_estimate_1}) agree very well. And
at the same time, we observe clear discrepancy between Eqs.~(\ref{eq:tau_erg_from_autocorrelation})
and (\ref{eq:ergodization_time_estimate_1}) for the one-dimensional
lattice, which indicates that the system has not ergodized on the
timescale covered by the Loschmidt echo. Non-ergodized fast growing
samples in $W(\Delta t)$ from Eq.~(\ref{eq:W_definition}) reach
a plateau significantly earlier than others: an indication of this
in Fig.~\ref{fig:Loschmidt_echo_for_dN_vs_LLE} is an early departure
of $W(\Delta t)$ from the linear growth regime. Overall, the ergodization
time decreases with the increasing lattice dimension, being the longest
in the one-dimensional case. Slow ergodization of one-dimensional
chains~(for Fermi-Pasta-Ulam, Klein-Gordon chains and also DGPE)
has been also noticed and investigated in Refs.~\cite{thudiyangal2017gross,danieli2017intermittent}.

In all three cases, we further observe that the values obtained from
Eq.~(\ref{eq:ergodization_time_rough_estimate_2}) give a satisfactory
approximation to Eq.~(\ref{eq:tau_erg_from_autocorrelation}).

We also performed the ergodicity test associated with relation~(\ref{eq:variance_of_G}).
The results are presented in Fig.~\ref{fig:SUPP_X_sqr_plateau},
where the ratio $\frac{\sigma_{G}^{2}(\Delta t)}{2\Delta t}$ is plotted
as a function of the echo time $\Delta t$. For the quickly ergodizing
two- and three-dimensional systems, the above ratio levels off rather
quickly around the expected value $\Lambda-\lambda_{\max}$, whereas
for the slow-ergodizing one-dimensional case it never reaches the
expected plateau.

\section{Relation to quantum systems}

Quantum-mechanical description of Loschmidt echoes involves out-of-time-order
correlators~\cite{fine2014absence,elsayed2015sensitivity,serbyn2017loschmidt,schmitt2017irreversible,hashimoto2017out,garttner2017measuring,rozenbaum2017lyapunov,schmitt2018semiclassical,tsuji2018out,garttner2018relating,hamazaki2018operator,cotler2018out, rammensee2018many, garcia2018chaos, kurchan2018quantum}.
We now illustrate that the parameter $\Lambda$ defined in the present
work from the relation 
\begin{equation}
\left\langle \left|\Delta\mathbf{X}(\Delta t)\right|\right\rangle \sim\exp(\Lambda\Delta t)\label{eq:LE_OTOC_lambda}
\end{equation}
also characterizes the growth of an OTOC in a quantum system, when
the system's constituents are describable quasi-classically. Following
Ref.~\cite{fine2014absence}, we observe that relation~(\ref{eq:LE_OTOC_lambda})
implies $\left\langle \left|\Delta\mathbf{X}(\Delta t)\right|^{2}\right\rangle \sim\exp(2\Lambda\Delta t)$.
In other words, 
\[
\Lambda=\frac{1}{2}\frac{d}{d(\Delta t)}\left[\lim_{\Delta t\to\infty;\left|\Delta\mathbf{X}(0)\right|\to0}\ln\left\langle \left|\Delta\mathbf{X}(\Delta t)\right|^{2}\right\rangle \right],
\]
where $\left\langle \left|\Delta\mathbf{X}(\Delta t)\right|^{2}\right\rangle =\left\langle \left(\mathbf{X}(\tau+\Delta t)-\mathbf{X}(\tau-\Delta t)\right)^{2}\right\rangle =\mbox{2\ensuremath{\left\langle \mathbf{X}^{2}\right\rangle }- \ensuremath{\left\langle \mathbf{X}(\tau+\Delta t)\mathbf{X}(\tau-\Delta t)\right\rangle }- \ensuremath{\left\langle \mathbf{X}(\tau-\Delta t)\mathbf{X}(\tau+\Delta t)\right\rangle }}$.
We focus here on the last two terms. They are equal classically, but
can become different when averaged quantum-mechanically. Both of them
become OTOCs in the quantum limit. Below, we show this for $\left\langle \mathbf{X}(\tau+\Delta t)\mathbf{X}(\tau-\Delta t)\right\rangle $. 

Let us consider a quantum system in equilibrium described at $t=0$
by Hamiltonian $\hat{\mathcal{H}}$. The density matrix of the system
is $\hat{\rho}_{0}\cong\exp\left(-\frac{\hat{\mathcal{H}}}{T}\right)$.
We are interested in the fluctuations of an observable quantity represented
by quantum operator $\hat{\mathbf{X}}$. Let us further assume that
the Hamiltonian changes sign at $t=\tau$, and, at the same moment
of time, the system components experience an infinitesimally small
random perturbation describable by quantum operator $\hat{R}$ (see
Ref.~\cite{fine2014absence} for a concrete example). As a result,
operator $\mathbf{\hat{X}}$ evolves as 

\begin{equation}
\mathbf{\hat{X}}(t)=\begin{cases}
\begin{array}{c}
e^{i\hat{\mathcal{H}}t}\hat{\mathbf{X}}(0)e^{-i\hat{\mathcal{H}}t},\\
e^{i\hat{\mathcal{H}}\tau}\hat{R}^{+}e^{-i\hat{\mathcal{H}}(t-\tau)}\hat{\mathbf{X}}(0)e^{i\hat{\mathcal{H}}(t-\tau)}\hat{R}e^{-i\hat{\mathcal{H}}\tau},
\end{array} & \begin{array}{c}
t<\tau\\
t>\tau
\end{array}\end{cases}.\label{eq:time_evolution_of_X}
\end{equation}
We now consider the quantum average $\mbox{\ensuremath{\left\langle \mathbf{\hat{X}}(\tau+\Delta t)\mathbf{\hat{X}}(\tau-\Delta t)\right\rangle \equiv\text{Tr}\left\{  \mathbf{\hat{X}}(\tau+\Delta t)\mathbf{\hat{X}}(\tau-\Delta t)\hat{\rho}_{0}\right\} } }$,
which, with the help of Eq.~(\ref{eq:time_evolution_of_X}) and a
simple manipulation, can be transformed into
\begin{equation}
\left\langle \mathbf{\hat{X}}(\tau+\Delta t)\mathbf{\hat{X}}(\tau-\Delta t)\right\rangle =\text{Tr}\left\{ \hat{R}^{+}\hat{\mathbf{X}}(-\Delta t)\hat{R}\hat{\mathbf{X}}(-\Delta t)\hat{\rho}_{0}\right\} .\label{eq:Loschmidt_echo_as_OTOC}
\end{equation}
Noting that $\hat{\mathbf{X}}$, as a physical observable, must be
describable by a Hermitian operator, i.e. $\mathbf{\hat{X}}^{+}(t)=\mathbf{\hat{X}}(t)$,
we rewrite Eq.~(\ref{eq:Loschmidt_echo_as_OTOC}) as $\left\langle \mathbf{\hat{X}}(\tau+\Delta t)\mathbf{\hat{X}}(\tau-\Delta t)\right\rangle =\left\langle \hat{R}^{+}\hat{\mathbf{X}}^{+}(-\Delta t)\hat{R}\hat{\mathbf{X}}(-\Delta t)\right\rangle $,
which is the standard form of OTOC. 

Finally, we note that the quantum counterpart of the maximum classical
Lyapunov exponent can be defined as

\[
\lambda_{\max}^{Q}=\frac{1}{2}\frac{d}{d(\Delta t)}\lim_{\Delta t\to\infty;\hat{R}\to\hat{\mathds{1}}}\text{Tr}\left\{ \hat{\rho}_{0}\ln\left(\mathbf{\hat{X}}^{2}(\tau+\Delta t)+\right.\right.
\]

\[
+\mathbf{\hat{X}}^{2}(\tau-\Delta t)-\mathbf{\hat{X}}(\tau+\Delta t)\mathbf{\hat{X}}(\tau-\Delta t)-
\]

\[
\left.\left.-\mathbf{\hat{X}}(\tau-\Delta t)\mathbf{\hat{X}}(\tau+\Delta t)\right)\right\} .
\]

It was proposed recently in Ref.~\cite{maldacena2016bound}, that
one can impose a temperature-dependent constraint on the exponential
growth rate $\Lambda$ of OTOCs (when the exponential growth regime
exists, which is not always the case~\cite{fine2014absence}). The
constraint on $\Lambda$, in turn, imposes a constraint on the largest
Lyapunov exponent $\lambda_{\max}$ for a quantum system. As follows
from the present work, as well as from Refs.~\cite{elsayed2015sensitivity,rozenbaum2017lyapunov,chavez2018quantum},
the value of $\Lambda$ is, in general, larger than $\lambda_{\max}$.
The interesting question then arises whether the difference between
$\Lambda$ and $\lambda_{\max}$ approaches zero as the number of
degrees of freedom in a system increases. Our findings indicate that,
for a lattice of a given dimension (1D, 2D and 3D), $\Lambda-\lambda_{\max}$
remains finite for rather large systems. Yet, this difference decreases
with the increase of the lattice dimension from 1D to 2D to 3D. It
is particularly small for the 3D lattice considered in this work,
which is consistent with the classical spin simulations for 3D lattices
done in Ref.~\cite{fine2014absence}, where the difference between
$\Lambda$ and $\lambda_{\max}$ was within the computational uncertainty
of the simulation and, hence, was overlooked. 

We further remark that the difference $\Lambda-\lambda_{\max}$ originates
from the fluctuations of Loschmidt echo amplitude, which is, as shown
in the present work, sensitive to ergodicity breakdown in classical
systems. The counterpart of this breakdown in quantum systems is the
transition from an ergodic to a many-body localized phase. It was
proposed in a related study~\cite{serbyn2017loschmidt}, that the
fluctuations of a Loschmidt echo in quantum systems are sensitive
to the many-body localization transition.

Finally, even though the primary agenda of the present article is
to characterize ergodicity in large systems close to the thermodynamic
limit, our method based on Loschmidt echoes should also be applicable
to a-few-body systems. When classical a-few-body systems exhibit the
breakdown of ergodicity, the ergodicity criterion proposed in Section~\ref{sec:Criterion-of-ergodicity}
should be sensitive to this. As far as a-few-body quantum systems
are concerned~\cite{percival1973regular,alhassid1992chaos,guhr1998random},
it is an interesting question how their energy level spacing statistics
is related to our ergodicity criterion in the classical limit. If
a quantum system exhibits the Wigner-Dyson statistics of energy-level
spacings in one energy range and does not exhibit it in the other
one, then the respective energy shells in the classical limit likely
change from ergodic to nonergodic. In such a case, the ergodicity
criterion of Section~\ref{sec:Criterion-of-ergodicity} can be used
to predict the level spacing statistics.

\section{Conclusion}

To summarize, we proposed a method of estimating ergodization time
of a chaotic many-particle system by monitoring equilibrium noise
before and after time reversal of dynamics, and validated it numerically
by simulations of the discrete Gross-Pitaevskii equation. We showed
that the difference between the largest Lyapunov exponent and the
growth rate of the classical counterpart of OTOCs is proportional
to the ergodization time of a system. We also introduced a related
test for the breakdown of ergodicity.

\section*{Acknowledgments}

We acknowledge discussions with S.~Flach and D.~Campbell. This work
was supported by a grant of the Russian Science Foundation (Project
No. 17-12-01587).

\appendix

\section{limits (5) and (6): independence of the observable $\mathbf{X}$\label{sec:from_phase_space_dist_to_G_and_W}}

If an experiment can track all phase-space coordinates of a system,
then it can obtain the largest Lyapunov exponent by identifying the
phase-space direction $\mathbf{\delta R}$ along which the growth
of a perturbation is the quickest, i.e. the eigenvector corresponding
to the largest local Lyapunov exponent. However, a realistic experiment
is limited to an observable $\mathbf{X}$. In such a case the eigenvector
is unlikely to belong to the subspace of the whole phase space that
contains $\mathbf{X}$, but it is overwhelmingly likely to have a
non-zero projection onto that subspace. This means that 
\begin{equation}
\Delta\mathbf{X}(\Delta t)=\Delta\mathbf{X}(0)\cos\alpha(\Delta t)e^{\int_{0}^{\Delta t}\lambda(t')dt'},\label{eq:def_angle_alpha}
\end{equation}
where $\alpha(\Delta t)$ is the angle between the eigenvector and
the direction corresponding to $\Delta\mathbf{X}(\Delta t)$ in the
many-dimensional phase space.

Here we consider the growth of the initial difference $\Delta\mathbf{X}(0)$
introduced by an imperfect time reversal, and justify the limits $\Delta t\to\infty$
for $G(\Delta t)$ in Eq.~(5) (cf. Ref.~\cite{fine2014absence})

\begin{equation}
G(\Delta t)\equiv\left\langle \ln\left|\Delta\mathbf{X}(\Delta t)\right|\right\rangle \xrightarrow[\Delta t\to\infty]{}\lambda_{\max}\Delta t\label{eq:G_definition-1}
\end{equation}
and for $W(\Delta t)$ in Eq.~(6)

\begin{equation}
W(\Delta t)\equiv\ln\left\langle \left|\Delta\mathbf{X}(\Delta t)\right|\right\rangle \xrightarrow[\Delta t\to\infty]{}\Lambda\Delta t.\label{eq:W_definition-1}
\end{equation}
We use Eq.~(\ref{eq:def_angle_alpha}) to express $G(\Delta t)$
as

\begin{equation}
G(\Delta t)=\left\langle \ln\left|\Delta\mathbf{X}(0)\right|+\ln\left|\cos\alpha(\Delta t)\right|+\ln e^{\int_{0}^{\Delta t}\lambda(t')dt'}\right\rangle ,\label{eq:G_angle_fluctuations}
\end{equation}
where the first term is constant, the second term remains limited
from above after ensemble averaging over initial conditions, and the
third term is the only one growing linearly with $\Delta t$. The
second term $\ln\left|\cos\alpha(\Delta t)\right|$ may appear problematic
for $\Delta t$ corresponding to $\left|\cos\alpha(\Delta t)\right|=0$.
However, this singularity is integrable: it vanishes after ensemble
averaging. Given the definition of $\lambda_{\max}$ from the main
text of the article, Eq.~(\ref{eq:G_angle_fluctuations}) implies
Eq.~(\ref{eq:G_definition-1}).

To prove the limit~(\ref{eq:W_definition-1}) for $W(\Delta t)$,
we assume that $\left|\cos\alpha(\Delta t)\right|$ is uncorrelated
with $e^{\int_{0}^{\Delta t}\lambda(t')dt'}$ and hence factorize
the average $\left\langle \left|\Delta\mathbf{X}(0)\cos\alpha(\Delta t)\right|e^{\int_{0}^{\Delta t}\lambda(t')dt'}\right\rangle \xrightarrow[\Delta t\to\infty]{}\left\langle \left|\Delta\mathbf{X}(0)\cos\alpha(\Delta t)\right|\right\rangle \cdot\left\langle e^{\int_{0}^{\Delta t}\lambda(t')dt'}\right\rangle $.
This assumption is, presumably, appropriate for almost any non-local
observable. It is supported by the extensive numerical experience,
e.g. Refs.~\cite{de2012largest,de2013lyapunov,fine2014absence,elsayed2015sensitivity},
showing that the eigenvectors corresponding to $\lambda_{\max}$ exhibit
rather erratic behavior. The above factorization leads to

\begin{equation}
W(\Delta t)=\ln\left\langle \left|\Delta\mathbf{X}(0)\cos\alpha(\Delta t)\right|\right\rangle +\ln\left\langle e^{\int_{0}^{\Delta t}\lambda(t')dt'}\right\rangle .\label{eq:W_angle_fluctuations}
\end{equation}
Given the definition of $\Lambda$, Eq.~(\ref{eq:W_angle_fluctuations})
implies Eq.~(\ref{eq:W_definition-1}).

\section{Derivation of Eq.~(8)\label{sec:Gaussian-random-variable}\label{sec:Calculation-of-the_correction}}

Here we derive Eq.~(8)

\begin{equation}
\left\langle e^{\int_{0}^{t}\delta\lambda(t')dt'}\right\rangle =e^{t\int_{0}^{\infty}\varphi(t')dt'},\label{eq:equation_for_derivation_in_SUPP-1}
\end{equation}
by a stochastic-noise method analogous to the one developed by Anderson
and Weiss~\cite{anderson1953exchange} in a different context, namely,
for the calculation of exchange-narrowed magnetic resonance linewidths.

We represent the left-hand side of Eq.~(\ref{eq:equation_for_derivation_in_SUPP-1})
as 

\begin{equation}
\left\langle e^{\int_{0}^{t}\delta\lambda(t')dt'}\right\rangle =\int dYP_{t}(Y)e^{Y},\label{eq:averaging_exponential_correction}
\end{equation}
where

\begin{equation}
Y(t)=\int_{0}^{t}\delta\lambda(t')dt'=\lim_{\delta t\to0}\delta t\sum_{t_{i}}\delta\lambda(t_{i}),\label{eq:Gaussianity_of_X}
\end{equation}
and $P_{t}(Y)$ is the probability distribution of $Y(t)$. We assume
that the system fluctuates near equilibrium, and, therefore, the process
$\delta\lambda(t)$ is stationary, i.e. its probability distribution
$p(\delta\lambda(t_{i}))$ is independent of $t_{i}$.

If $\delta\lambda(t)$ is a Gaussian random variable, then $Y$ is
also a Gaussian random variable for all times, i.e. $P_{t}(Y)$ is
Gaussian. If $p(\delta\lambda)$ is not Gaussian, but the variable
$\delta\lambda(t)$ has a finite memory time $\tau_{erg}$, then $P_{t}(Y)$
still becomes Gaussian for $t\gg\tau_{erg}$ (consequence of the central
limit theorem). 

Assuming Gaussianity, $P_{t}(Y)\equiv\left(2\pi\left\langle Y(t)^{2}\right\rangle \right)^{-\frac{1}{2}}\exp\left(-\frac{Y^{2}}{2\left\langle Y(t)^{2}\right\rangle }\right)$.
Eq.~(\ref{eq:averaging_exponential_correction}) now reads:

\begin{equation}
\left\langle e^{\int_{0}^{t}\delta\lambda(t')dt'}\right\rangle =\left(2\pi\left\langle Y^{2}\right\rangle \right)^{-\frac{1}{2}}\int dYe^{-\frac{Y^{2}}{2\left\langle Y^{2}\right\rangle }+Y}=e^{\frac{\left\langle Y^{2}\right\rangle }{2}}.\label{eq:via_gaussianity}
\end{equation}
We calculate the variance of $Y$ as

\begin{equation}
\left\langle Y^{2}\right\rangle =\left\langle \left[\int_{0}^{t}\delta\lambda(t')dt'\right]^{2}\right\rangle =\int_{0}^{t}dt'\int_{0}^{t}dt''\left\langle \delta\lambda(t')\delta\lambda(t'')\right\rangle .\label{eq:variance_of_Y_0}
\end{equation}
Since $\delta\lambda(t)$ is assumed to be stationary: $\left\langle \delta\lambda(t')\delta\lambda(t'')\right\rangle =\left\langle \delta\lambda(0)\delta\lambda(t''-t')\right\rangle \equiv\varphi(t''-t')$,
and Eq.~(\ref{eq:variance_of_Y_0}) becomes

\begin{equation}
\left\langle Y^{2}(t)\right\rangle =\int_{0}^{t}dt'\int_{-t'}^{t-t'}\varphi(t'')dt''=\int_{0}^{t}dt'g(t'),\label{eq:variance_x2_Loschmidt_echo}
\end{equation}
where $g(t')=\int_{-t'}^{t-t'}\varphi(t'')dt''$. The dynamics is
time-reversible, thus $\varphi(-t')=\varphi(t')$, and $\dot{g}(t')=-\varphi(t-t')+\varphi(-t')=-\varphi(t-t')+\varphi(t')$.
Integrating Eq.~(\ref{eq:variance_x2_Loschmidt_echo}) by parts leads
to

\[
\left\langle Y^{2}(t)\right\rangle =t\cdot g(t)-\int_{0}^{t}dt'\cdot t'\dot{g}(t')=
\]

\[
=t\int_{-t}^{0}\varphi(t'')dt''-\int_{0}^{t}dt'\cdot t'\varphi(t')+\int_{0}^{t}dt'\cdot t'\varphi(t-t')=
\]
\[
=t\int_{0}^{t}\varphi(t')dt'-\int_{0}^{t}dt'\cdot t'\varphi(t')+\int_{0}^{t}dt'\cdot(t-t')\varphi(t')=
\]

\begin{equation}
=2\int_{0}^{t}dt'(t-t')\varphi(t').\label{eq:final_dispersion_value}
\end{equation}
We substitute Eq.~(\ref{eq:final_dispersion_value}) into Eq.~(\ref{eq:via_gaussianity}),
and finally obtain

\begin{equation}
\left\langle e^{\int_{0}^{t}\delta\lambda(t')dt'}\right\rangle =e^{\int_{0}^{t}dt'(t-t')\varphi(t')}.\label{eq:final_result_for_the_correction}
\end{equation}
This integral converges if $\varphi(t)$ decays faster than $\frac{1}{t^{2}}$.
In such a case, for $t\to\infty$

\begin{equation}
\left\langle e^{\int_{0}^{t}\delta\lambda(t)dt}\right\rangle =Ce^{t\int_{0}^{t}dt'\varphi(t')},\label{eq:average_via_autocorr}
\end{equation}
where $C=\exp\left(-\int_{0}^{\infty}dt'\cdot t'\varphi(t')\right)$.

\bibliographystyle{apsrev4-1}
\bibliography{references}

\end{document}